\documentclass[10pt,a4paper]{article}
\usepackage{url}
\usepackage{graphicx}
\usepackage[left=3cm,top=2cm,right=1cm,bottom=2cm,nohead]{geometry}

\newcommand\rmd{\mathrm{d}}
\newcommand\erfc{\mathrm{erfc}}

\begin{document}

\author{Vygintas Gontis\thanks{Vilnius University, Institute of Theoretical Physics and Astronomy, vygintas@gontis.eu}, Aleksejus Kononovicius, Stefan Reimann}

\title{The class of nonlinear stochastic models as a background for the bursty behavior in financial markets}

\maketitle

\begin{abstract}
We investigate behavior of the continuous stochastic signals above some threshold, bursts, when the exponent of multiplicativity is higher than one. Earlier we have proposed a general nonlinear stochastic model applicable for the modeling of absolute return and trading activity in financial markets which can be transformed into Bessel process with known first hitting (first passage) time statistics. Using these results we derive PDF of burst duration for the proposed model. We confirm derived analytical expressions by numerical evaluation and discuss bursty behavior of return in financial markets in the framework of modeling by nonlinear SDE.
\end{abstract}

\section{Introduction}
Most econometric analysis of financial markets based on the
various versions of
stochastic differential equations (SDEs) have
been limited to the case when the exponent of noise
multiplicativity $\eta$ is lower than $1$ \cite{jeanblanc2009}.
This is related to the general econometric assumptions for the
asset price process and existence of unique Martingale measure
\cite{Davydov2001}. Nevertheless, Chan et al. \cite{Chan1992} have
shown, comparing varying econometric models of short-interest
rate, that the models allowing multiplicativity of $\eta > 1$ capture
the volatility changes of short-term interest rates better than
those with $\eta < 1$.

We have introduced a class of non-linear stochastic differential
equations (SDEs) providing time series with power-law statistics,
and most notably reproducing $1/f$ spectral density,
\cite{Gontis2004PhysA343,Kaulakys2005PhysRev,Ruseckas2011EPL}. The
general expression of the proposed class of Ito SDEs is
\begin{equation}
\rmd
x=\left(\eta-\frac{\lambda}{2}\right)x^{2\eta-1} \rmd t_s
+ x^{\eta}\rmd W_s .\label{eq:sde}
\end{equation}
Here $x$ is the stochastic process exhibiting power-law
statistics, $\eta$ is the power-law exponent of the
multiplicative noise, $\lambda$ is the exponent of power-law probability density function (PDF), and $W$ is a standard Wiener process
(the Brownian motion). Note that SDE~(\ref{eq:sde}) was used in the modeling of trading activity and absolute return of financial markets \cite{Gontis2010Sciyo} and
is defined in the dimensionless scaled time, $t_s$, we denote the scaling constant as
$\sigma_t^2$ and will use the relation to the real time $t$ of financial markets as $t_s = \sigma_t^2 t$. Empirical value $\sigma_t^2=1/6 \cdot 10^{-5} (\mathrm{s}^{-1})$ appropriate for the markets absolute return model was defined in \cite{Gontis2010PhysA}.

Directly from the SDE~(\ref{eq:sde})
follows that the stationary probability density function (PDF) of
this stochastic process is power-law, $P_{0}(x)\sim x^{-\lambda}$,
with the exponent $\lambda$ \cite{gardiner1997}. While in
Refs.~\cite{Kaulakys2006PhysA} and later more precisely in
\cite{Ruseckas2010} it was shown that SDE~(\ref{eq:sde}) provides
time series with power-law spectral density
\begin{equation}
S(f)\sim \frac{1}{f^{\beta}} ,
\qquad\beta=1+\frac{\lambda-3}{2(\eta-1)} .
\end{equation}
Note that exponent of spectral density, $\beta$, is defined only for $\eta \neq 1$. In case of
$\eta =1$ the SDE~(\ref{eq:sde}) becomes identical to the geometric
Brownian motion and in case $\lambda =3$ one gets $1/f$ noise. Do not be confused by cumulative inverse cubic law defined for return in \cite{Gabaix2006}, where $\lambda=4$.

Empirical data from the financial markets confirm the choice of
equations~(\ref{eq:sde}) with $\eta
>1$ \cite{Gontis2010Sciyo} while modeling trading activity and absolute return.
The case of $\eta > 1$ represents the higher positive
signal-to-noise feedback and leads to the faster than exponential
changes, growth or descent, observed in the time series
\cite{ReimmanPhysA2011}.

Our modeling of financial market variables using the non-linear
stochastic differential equations is based on the empirical
analysis and the power-law statistics of proposed equations
\cite{Gontis2010Sciyo}. Providing microscopic, agent-based,
reasoning for the proposed equations seems to be a formidable task
for such complex system. Apparently, the development of
macroscopic descriptions for the well established agent based
models would be more consistent approach towards the understanding
of the micro and macro correspondence. For such analysis one
should select simple agent based models with established or
expected macroscopic description. First of all we expect that
herding, contagion and cascading behavior of agents should lead to
the macroscopic description with the non-linear stochastic
differential equations.

The convenience of this approach is already confirmed by some
recent publications considering this problem
\cite{Daniunas2011,Kononovicius2012PhysA,Ruseckas2011EPL}.
Kirman's ant colony model \cite{Kirman1993} is an agent-based
model, which explains the importance of herding. As human crowd
behavior ideologically is very similar, the ant colony model
actually was built as a general framework for financial market
modeling \cite{Alfarano2005,Alfarano2008,Kirman1993}. In
\cite{Daniunas2011,Kononovicius2012PhysA,Ruseckas2011EPL} we
follow the works by Alfarano et al.
\cite{Alfarano2005,Alfarano2008} to introduce model modifications,
which allow us to obtain agent-based models for
the financial markets exhibiting a macroscopic
description in terms of SDEs with $\eta > 1$.

In this contribution we demonstrate that the general class of
SDEs~(\ref{eq:sde}) can be transformed to the Bessel process, which
represents a special family of diffusion models applicable in
econometric analysis \cite{jeanblanc2009}. This
allows us to derive explicit form of burst statistics generated by
the SDE~(\ref{eq:sde}). Exponent of multiplicativity $\eta$ is a
key parameter of defined statistics. We also present analysis of
empirical data providing an evidence that
burst statistics of return in financial markets can be modeled by non-linear
stochastic differential equations with multiplicativity as high as $\eta=\frac{5}{2}$.

\section{Statistics of stochastic bursty time series}

Looking for the best form of SDE applicable to the modeling of complex systems one has to use
the most distinct features of the observed behavior. Spikes, bubbles or bursts of the observed signals is one of the characteristic
features of complex systems, as well as 1/f noise. Well defined statistics of bursty behavior can serve as
an additional source of information about the system. It is natural to expect that parameter of multiplicativity $\eta$
in Eq. (\ref{eq:sde}) is responsible for the bursty behavior of defined stochastic signal. Therefore described statistics of bursty behavior would help establish relation between model parameters and empirical data. Although we concentrate here more on the return not price dynamics, such characteristics of bursts as duration and peak value should be
indispensable in risk valuation process.

\begin{figure}
\centering
\includegraphics[width=0.48\textwidth]{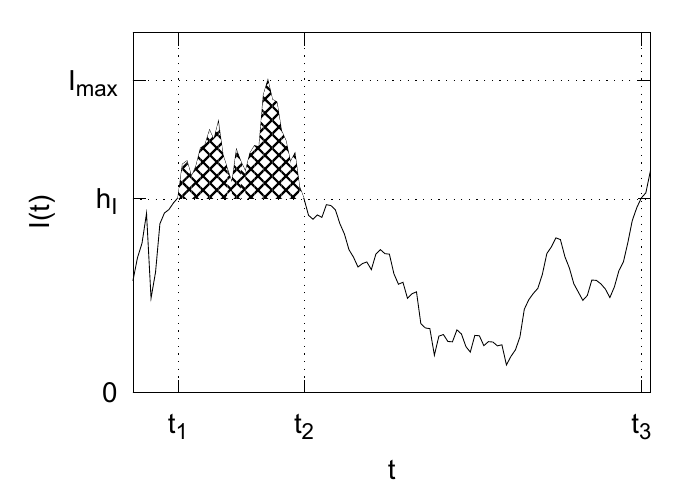}
\caption{Time series exhibiting bursty behavior, $I(t)$. Here $h_I$ is
threshold value, above which bursts are detected, $t_i$ is the
three visible threshold passage events, $I_{max}$ is the
highlighted burst's peak value. The other relevant statistical properties are defined in respect to those shown in this figure: $T= t_2 - t_1$, $\theta = t_3 - t_2$, $\tau = T+ \theta = t_3 - t_1$.} \label{fig:burstExample}
\end{figure}

We define a burst as a part of time series lying above certain
threshold, $h$. In Fig. \ref{fig:burstExample} we have presented
an example burst of the simple bursty time series, $I(t)$ (hence
the threshold is denoted as $h_I$). Evidently a burst
itself can be characterized by its duration, $T = t_2 - t_1$, peak
value, $I_{max}$, and burst size, defined as the area above the
threshold yet bellow time series (highlighted), $S$. One can also introduce
inter-burst, $\theta = t_3 - t_2$, and waiting, $\tau = t_3 -
t_1$, times to be able to fully grasp the statistical features of bursty behavior.
We will derive the explicit form PDF of burst duration $T$, generated by Eq.~(\ref{eq:sde}), and will analyze numerically correspondence with empirical data of returns and relations to the other statistical characteristics of bursty behavior.

There is a well established hitting (passage) time framework, which is of
high interest in both mathematical finance \cite{jeanblanc2009}
and physics \cite{gardiner1997,redner2001}. This
framework can be also applied in the description of burst duration PDF.
Actually first hitting time of the stochastic process starting infinitesimally near above the
hitting threshold is the same as burst duration. It is so as the first
passage of threshold ends the burst by its definition.

\subsection{Numerical and empirical definitions of bursty behaviour statistics}

In this contribution we consider stochastic model driven by SDE~(\ref{eq:sde})
and empirical time series of absolute return. Let us present a brief
discussion on how we deal with the numerical and empirical time series.

SDE~(\ref{eq:sde}) serves as a most simple definition of very wide class of stochastic processes with power law statistics for high values
of the signal intensity, $x$. Any real system with corresponding power law statistics of high values has to be restricted from the side of small values. One can consider reflecting boundary condition at the point $x=0$ as the most simple case. Nevertheless, the more general approach can be implemented by wide choice of compressed exponential restrictions, which can be introduced by additional term in SDE~(\ref{eq:sde})
\begin{equation}
\rmd
x=\left(\eta-\frac{\lambda}{2}+\frac{m}{2}\frac{x_{min}^m}{x^m}\right)x^{2\eta-1} \rmd t_s
+ x^{\eta}\rmd W_s .\label{eq:sdem}
\end{equation}
Here $x_{min}$ denotes the diffusion restriction point and $m$ defines how sharp restriction is. Steady state distribution of $x$ is as follows
\begin{equation}
P_0(x)=\frac{m x_{min}^{\lambda-1}}{\Gamma[\frac{\lambda-1}{m}]} \exp\left[-\left(\frac{x_{min}}{x}\right)^m \right] x^{-\lambda}
\end{equation}

As we analyze bursts, behavior in the region of high values of $x$, it is reasonable to assume $x_{min}=1$ and $m=2$. With such assumption we numerically solve (\ref{eq:sdem})
by evaluating the difference equations:
\begin{eqnarray}
& x_{i+1} = x_i + \kappa^2 \left(\eta - \frac{\lambda}{2} + \frac{1}{x_i^2} \right) x_i + \kappa \sqrt{x_i} \zeta_i , \label{eq:differenceX} \\
& t_{s,\,i+1} = t_i + \frac{\kappa^2}{x^{2 \eta -2}} , \label{eq:differenceT}
\end{eqnarray}
here $\zeta_i$ is a normally distributed random variable with unit variance and
zero mean, $\kappa$ is a numerical precision parameter (it should be significantly
smaller than 1). The above difference equations follow from the
Euler-Maruyama method for the numerical solution of the stochastic differential equations
\cite{Kloeden1999Springer} with variable time steps. Looking for the higher
precision one may alternatively choose Milstein
method \cite{Kloeden1999Springer}.

We iterate through the (\ref{eq:differenceX}) and (\ref{eq:differenceT})
until we obtain time series with set amount of bursts. The suitable  precision was achieved from time series with $10^5$ bursts.
We choose threshold value $h_x=2$ to ensure sufficient number of burst and $\lambda=4$ as well defined empirical exponent of return distribution \cite{Gabaix2006}, in our all numerical calculations.

Our empirical data set includes all trades made on NYSE, which were made from
January, 2005 to March, 2007 and involved 24 different stocks, ABT, ADM, BMY,
C, CVX, DOW, FNM, GE, GM, HD, IBM, JNJ, JPM, KO, LLY, MMM, MO, MOT, MRK,
SLE, PFE, T, WMT, XOM.  We have shown in \cite{Gontis2010Sciyo}
that the more sophisticated versions of (\ref{eq:sdem}) may be well used to
model absolute return and trading activity of different stocks from NYSE and Vilnius
Stock Exchange. In this approach the normalized absolute return time series of different stocks served as
independent realizations of the stochastic process driven by the same nonlinear SDE with the same set of parameters.
Such universal nature of stock market return behavior is hidden under secondary high frequency stochastic process with the $q$-Gaussian PDF.

The double stochastic nature of the return makes analysis of the bursty behavior in empirical stock data as a more complex task.
One minute return time series are very noisy, see \cite{Gontis2010Sciyo} for more details. This noise of normalized time series can be diminished by using moving average filter. We have used one hour window moving average filter on the empirical data of one minute absolute returns. Time series of all stocks were used to define statistics of burst durations, peak values and sizes.

\subsection{Obtaining probability density function of burst durations using the first hitting times of Bessel process}

There are few simple and highly applicable models for which hitting
times statistics are known. These models include, but are not limited to, Brownian motion,
geometric Brownian motion and Bessel process \cite{jeanblanc2009}. The Bessel process,
\begin{equation}
\rmd R = \frac{N-1}{2} \frac{\rmd t_s}{R} + \rmd W_s , \label{eq:BesselProcSDE}
\end{equation}
is one of the most interesting as some prominent mathematical
finance models can be transformed to a similar form. Bessel
process is also interesting in Physics since it describes the
evolution of the Euclidean norm, $R = | \vec R |$, of the $N$-dimensional
Brownian particle. The widely used parameter $\nu$,
\begin{equation}
\nu = \frac{N}{2} -1 ,
\end{equation}
is known as the index of the Bessel process related to the $N$-dimensional
Brownian diffusion. Note that for $N>1$, or alternatively $\nu>-0.5$,
$R$ tends to diverge towards infinity.

One can reduce (\ref{eq:sde}) to Bessel process by using Lamperti transformation,
$\ell:x \mapsto y(x)$. The exact form of transformation can be obtained by requiring that
\begin{equation}
x^\eta \partial_x y(x) = \pm 1 .
\end{equation}
Note that this requirement follows from the the Ito variable substitution formula \cite{gardiner1997}.
Thus the Lamperti transformation has the following form
\begin{equation}
\ell:x \mapsto y(x) = \frac{1}{(\eta-1) x^{\eta-1}} . \label{eq:varTrans}
\end{equation}
In such case the stochastic differential equation (\ref{eq:sde}) can be transformed
to the Bessel process,
\begin{equation}
\rmd y = \left( \nu + \frac{1}{2} \right) \frac{\rmd
t_s}{y} + \rmd W_s , \label{eq:BesselTransSDE}
\end{equation}
with index $ \nu = \frac{\lambda - 2 \eta + 1}{2 (\eta -1)}$. The dimension of the
obtained Bessel process is given by $ N = 2 (\nu+1)= \frac{\lambda-1}{\eta-1} $.

Let us assume that a burst starts at time $t_0$, thus $x_0 = x(t_0)$
exceeds the threshold, $h_x$, by some small amount.
The burst lasts until $x(t)$ crosses back $h_x$, from the above.
Equivalently, in the terms of Bessel process the burst lasts until at
a certain time, $t$, the $y$ process crosses the boundary $h_y:=\ell(h_x)$
from the below, while the starting position, $y_0 =y(t_0)$, in terms of Bessel process also
lies below the threshold, $y_0 = \ell(x_0)< h_y$. We see that
the burst durations of the stochastic process driven by SDE (\ref{eq:sde})
is related to the well known first passage times of the Bessel process.

By choosing $y_0$ arbitrarily close yet below $h_y$, we can obtain an estimate for the burst duration, $T$, in terms of first passage times of the Bessel process, $\tau^{(\nu)}_{y_0,h_y}$:
\begin{equation}
T = \tau^{(\nu)}_{y_0,h_y} = \inf_{t>t_0} \Big\{ t , \: y(t) \geq h_y \Big\} , \quad 0 < h_y-y_0 < \epsilon ,
\end{equation}
where $\epsilon$ is an arbitrary small positive constant. As given in \cite{Borodin2002Birkhauser}, the following holds for $0 < y_0 < h_y$:
\begin{equation}
\rho^{(\nu)}_{y_0,h_y}(t) = \frac{h_y^{\nu-2}}{y_0^\nu} \sum_{k=1}^{\infty} \frac{j_{\nu,k} J_\nu\left(\frac{y_0}{h_y} j_{\nu,k}\right)}{J_{\nu+1}(j_{\nu,k})} \exp\left(- \frac{j_{\nu,k}^2}{2 h_y^2} t\right) ,
\end{equation}
where $\rho_{y_0,h_y}^{(\nu)}(t)$ is a probability density function of the first passage times at level $h_y$ of Bessel process with index $\nu$ starting from $y_0$, $J_\nu$ is a Bessel function of the first kind of order $\nu$ and
$j_{\nu,k}$ is a $k$-th zero of $J_\nu$.

\begin{figure}
\centering
\includegraphics[width=0.48\textwidth]{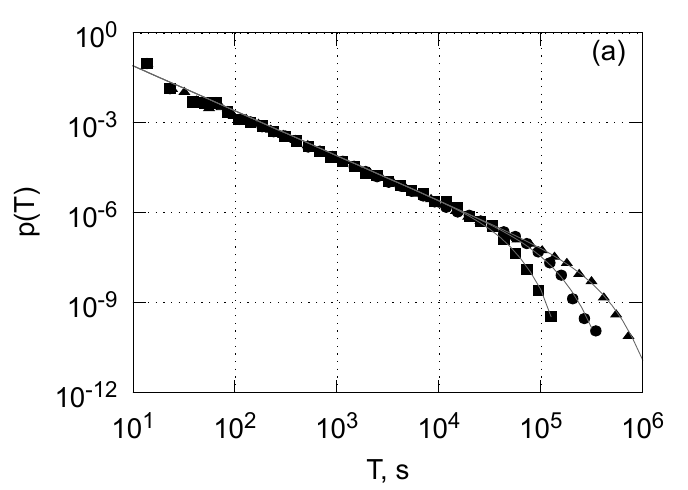}
\includegraphics[width=0.48\textwidth]{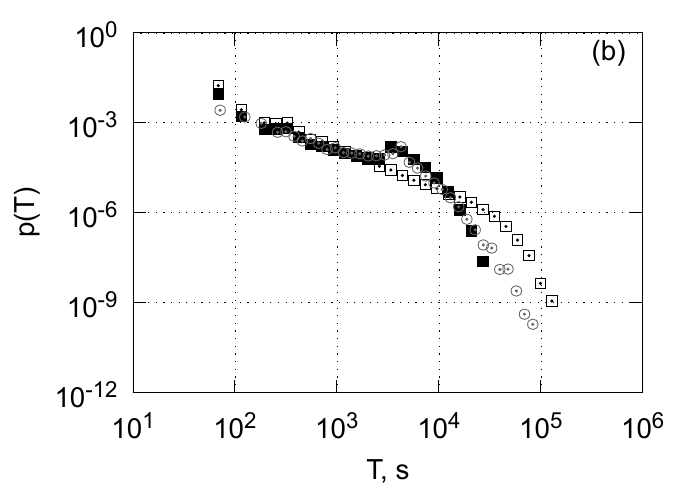}
\caption{ Simple SDE (a) and empirical versus the double stochastic model (b) PDF of the burst
durations, $h_x=2$. In sub-figure (a) numerical data is represented
by the filled shapes, while fits, (\ref{eq:durationApprox}), are represented by gray curves.
Simple SDE, (\ref{eq:sde}), parameters were set as follows: $\sigma_t^2 = 1/6 \cdot 10^{-5} \mathrm{s}^{-1}$ (in all three cases), $\lambda=4$ (in all three cases), $\eta = 2.5$ (squares, $\nu =0$), $\eta = 2$ (circles, $\nu =0.5$) and $\eta=1.5$ (triangles, $\nu =2$). Empirical data (gray empty circles in the sub-figure (b) is plotted versus the numerical data from the complex SDE, (\ref{eq:return2}), (black empty squares) and from the double stochastic model (black filled squares). Complex SDE, (\ref{eq:return2}), parameters were set as follows: $\sigma_t^2 = 1/6 \cdot 10^{-5} (\mathrm{s}^{-1})$, $\eta=2.5$, $\lambda=3.6$, $\epsilon=0.017$, $x_{max}=10^3$. Double stochastic model uses the same parameters as complex SDE and the additional parameters, which were set as follows: $\bar{r}_0=0.4$, $\lambda_2=5$.}
\label{fig:burstDuration}
\end{figure}

We have to replace $\rho^{(\nu)}_{y_0,h_y}(t)$ by density function regarding $h_y$ to avoid trivial convergence of $\rho^{(\nu)}_{y_0,h_y}(t)$ to zero, when $y_0 \rightarrow h_y$. This is achieved introducing PDF $p_{h_y}^{(\nu)}(t)$ as a probability density function of burst duration
\begin{equation}
p_{h_y}^{(\nu)}(t) = \lim_{y_0 \rightarrow h_y} \frac{\rho^{(\nu)}_{y_0,h_y}(t)}{h_y-y_0} , \label{eq:durationPrecise}
\end{equation}
where the threshold is set at level $h_y$ and $\nu$ is related to the model parameters
(see discussion on the transformation to Bessel process). To evaluate the limit we have to expand
$J_\nu\left(\frac{y_0}{h_y} j_{\nu,k}\right)$ near $\frac{y_0}{h_y} = 1$:
\begin{eqnarray}
& J_\nu\left(\frac{y_0}{h_y} j_{\nu,k}\right) \approx J_\nu( j_{\nu,k}) - \left(1-\frac{y_0}{h_y}\right) \left[\nu
J_\nu( j_{\nu,k}) - j_{\nu,k} J_{1+\nu}(j_{\nu,k}) \right] = \nonumber\\
& = (1-\frac{y_0}{h_y}) j_{\nu,k} J_{1+\nu}(j_{\nu,k}).
\end{eqnarray}
By using this expansion we can rewrite (\ref{eq:durationPrecise}) as:
\begin{equation}
p_{h_y}^{(\nu)}(t) = C_1 \sum_{k=1}^{\infty} j_{\nu,k}^2
\exp\left(- \frac{j_{\nu,k}^2}{2 h_y^2} t\right) ,
\label{eq:duration}
\end{equation}
here $C_1$ is a normalization constant. Since $j_{\nu,k}$ are almost equally
spaced \cite{Abramowitz1972}, we can replace the sum by integration
\begin{eqnarray}
& p_{h_y}^{(\nu)}(t) \approx C_2 \int_{j_{\nu,1}}^{\infty} x^2 \exp\left(-\frac{x^2 t}{2 h_y^2} \right)
\rmd x = \nonumber\\
& =C_2 \left[\frac{h_y^2 j_{\nu,1} \exp\left(-\frac{j_{\nu,1}^2 t}{2 h_y^2}\right)}{t}+ \sqrt{\frac{\pi}{2}} \frac{h_y^3
\erfc\left(\frac{j_{\nu,1} \sqrt{t}}{\sqrt{2} h_y}\right)}{t^{3/2}} \right] .
\label{eq:durationApprox}
\end{eqnarray}
Note that derived PDF in the form of Eqs. (\ref{eq:duration}) or (\ref{eq:durationApprox}) diverge, when $t$ approaches zero. 
Therefore normalization constants $C_1$ and $C_2$ can be defined if some minimum value of $t$ is supposed.
Accuracy of numerical calculations or minimum intertrade time can be considered as possible choices.
From the above follows that the distribution of burst durations of SDE
(\ref{eq:sde}) can be approximated by a power-law with exponential
cut-off according to
\begin{eqnarray}
& p_{h_y}^{(\nu)}(t) \sim t^{-3/2} , \quad \textrm{when}\quad t \ll \frac{2 h_y^2}{j_{\nu,1}^2},\\
& p_{h_y}^{(\nu)}(t) \sim \frac{\exp\left(-\frac{j_{\nu,1}^2 t}{2 h_y^2}\right)}{t} , \quad
\textrm{when}\quad  t \gg \frac{2 h_y^2}{j_{\nu,1}^2}
\end{eqnarray}
This result is in agreement with a general property of one dimensional diffusion processes
presented in \cite{redner2001}, namely that the asymptotic behavior of first hitting times
is a power-law $t^{-3/2}$ irrespectively of the nature of stochastic
process or the actual form of Langevin and Fokker-Plank equations.
The exponential cutoff for longer burst durations is caused by the direction preference,
note the positive drift term in case of $N>1$, or alternatively $\nu>-0.5$, of Bessel
processes. Numerical solutions of the SDE (\ref{eq:sde}) confirm the derived probability
density function, (\ref{eq:durationApprox}), of the burst duration, $T$, see Fig.
\ref{fig:burstDuration} (a).

Empirical data, as shown in Fig. \ref{fig:burstDuration} (b), also has similar asymptotic behavior
for short and long burst durations, though the fitting using (\ref{eq:durationApprox}) would be inconsistent
for the intermediate burst durations, note the cusp. There are a numerous reasons for
this. Firstly, we were unable to remove intra-day pattern from
the empirical time series. But the main reason is that in order to reproduce the correct shape
of the empirical probability density function for the intermediate values of $T$ one must use
the double stochastic model, driven by a more sophisticated version of the SDE (\ref{eq:sde}).
In Fig. \ref{fig:burstDuration} (b) we demonstrate pretty good agreement of the probability
density function of $T$ retrieved from the empirical NYSE time series and a more sophisticated
double stochastic return model discussed in Appendix A.

\subsection{Power law interdependencies between burst related variables}

\begin{figure}
\centering
\includegraphics[width=0.48\textwidth]{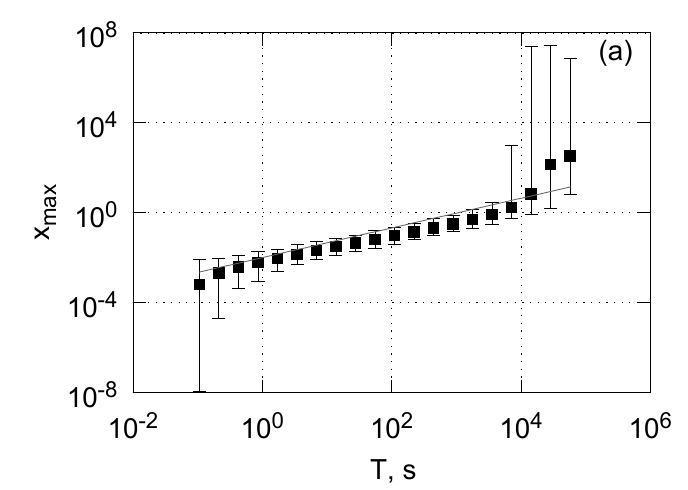}
\includegraphics[width=0.48\textwidth]{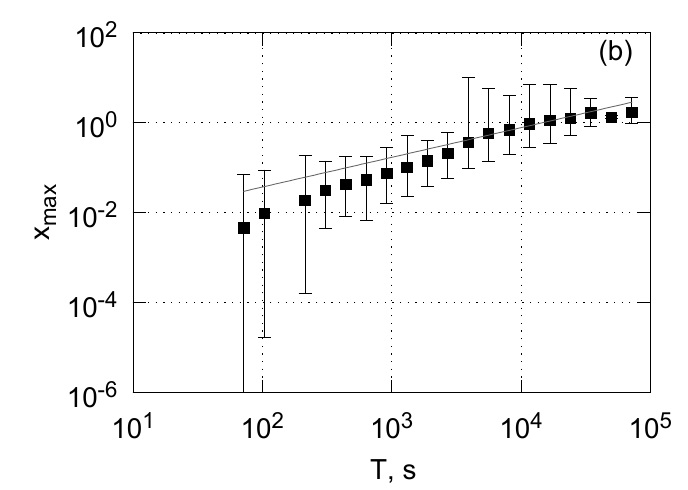}
\caption{The scatter plot of the numerical (a) and empirical (b) burst peak values vs burst durations, $h_x=2$. In both sub-figures filled squares represent mean values and error bars - variability, while power-law fits, $\alpha = \frac{2}{3}$, are represented by gray curves. Model, (\ref{eq:sde}), parameters were set as follows:  $\sigma_t^2 = 1/6 \cdot 10^{-5} \mathrm{s}^{-1}$, $\eta=2$, $\lambda=4$ ($\nu=0$).}
\label{fig:tmax}
\end{figure}

\begin{figure}
\centering
\includegraphics[width=0.48\textwidth]{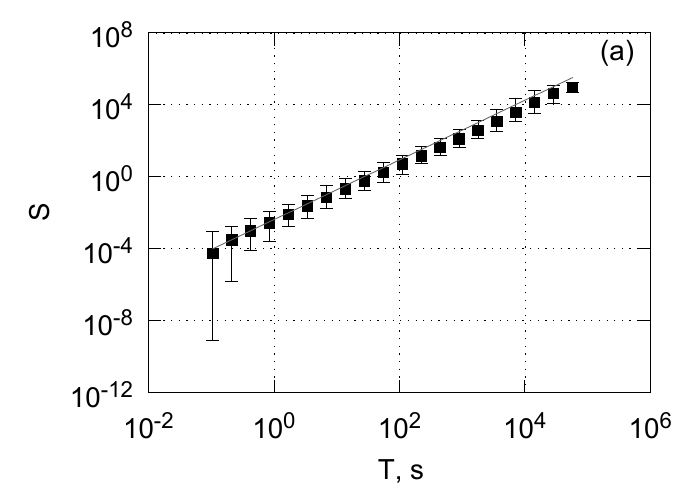}
\includegraphics[width=0.48\textwidth]{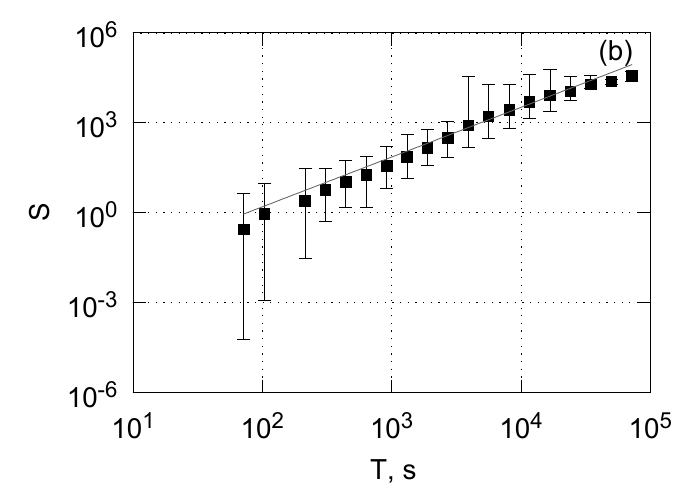}
\caption{The scatter plot of the numerical (a) and empirical (b) burst size  vs burst durations, $h_x=2$. In both sub-figures filled squares represent mean values and error bars - variability, while power-law fits, $\alpha=\frac{5}{3}$, are represented by gray curves. Model, (\ref{eq:sde}), parameters are the same as in Fig. \ref{fig:tmax}.}
\label{fig:tsize}
\end{figure}

\begin{figure}
\centering
\includegraphics[width=0.48\textwidth]{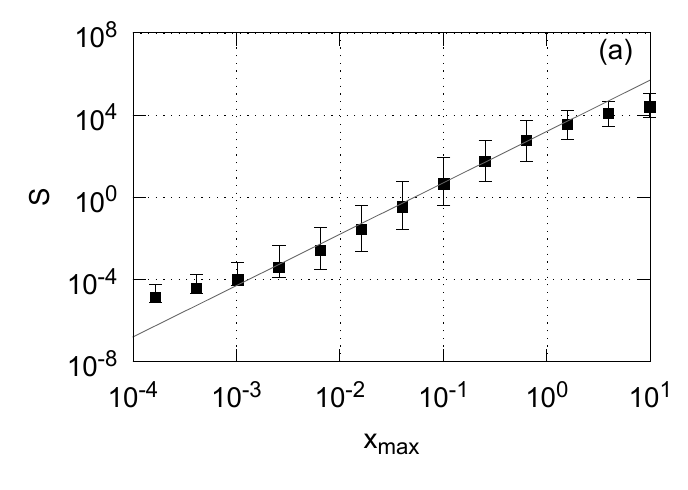}
\includegraphics[width=0.48\textwidth]{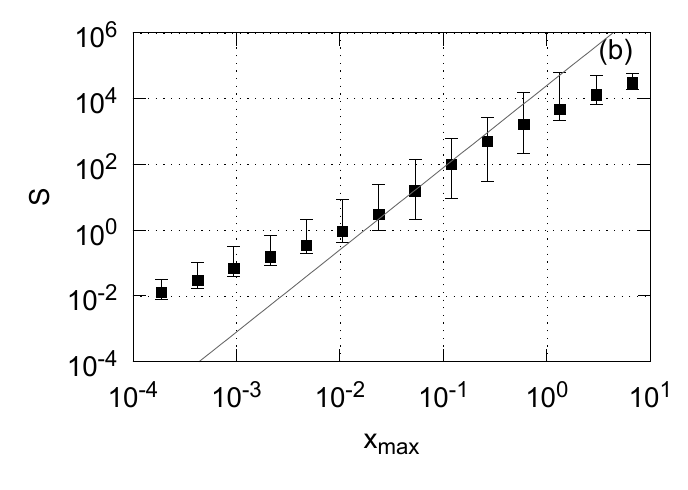}
\caption{The scatter plot of the numerical (a) and empirical (b) burst size vs burst peak values, $h_x=2$. In both sub-figures filled squares represent mean values and error bars - variability, while power-law fits, $\alpha=\frac{5}{2}$ are represented by gray curves. Model, (\ref{eq:sde}), parameters are the same as in Fig. \ref{fig:tmax}.}
\label{fig:maxsize}
\end{figure}

Another interesting feature of both numerical results and empirical data is
that burst duration, peak value and burst size are correlated or even interdependent, $(T,x_{max},S)$.
The scatter plots of these variables also reveal power-law asymptotic behavior
- $x_{max} \propto T^{\frac{2}{3}}$ (see Fig. \ref{fig:tmax}) and $S \propto T^{\frac{5}{3}}$
(see Fig. \ref{fig:tsize}). Consequently, as follows from these relations
$S \propto (x_{max}^{\frac{2}{3}})^\frac{5}{3} \propto x_{max}^{\frac{5}{2}}$ (see Fig. \ref{fig:maxsize}). From SDE (\ref{eq:sde}) numerically defined asymptotic relations
are shown to hold for both empirical data and numerical results, though the theoretical power-law holds only in narrow region of empirical data.
The interdependence between these variables suggest that the geometry
of burst remains qualitatively the same in the fitted $(T,x_{max},S)$ regions. Theoretical consideration of these relations in the region of $T$ PDF exponential cut-of  will be continued.

\section{Conclusions}
Starting from the general expression of nonlinear SDE (\ref{eq:sde}) and well-known PDF expression of first hitting time for Bessel process we have derived PDF of burst duration (\ref{eq:duration}) and its approximation (\ref{eq:durationApprox}) for the nonlinear stochastic process driven by (\ref{eq:sde}).
The understanding of burst statistics is needed to enable efficient modeling of trading activity and return in financial markets based on the various versions of SDE (\ref{eq:sde}) \cite{Gontis2010Sciyo}. We do expect that proposed class of SDE is applicable in the modeling of other complex systems and defined statistics of bursts can be helpful in risk analyzes.
We have also confirmed analytical results of burst statistics  by numerical calculations of SDE (\ref{eq:sde}), and compared it to the empirical data of return in financial markets.
The obtained results encourage us to continue research of burst statistics in the financial markets seeking to develop more sophisticated versions of stochastic models.
Most importantly we find that burst related statistical properties are related to the key parameter - exponent of stochastic multiplicativity $\eta$ characterizing the dynamics of risk in financial markets. Both numerical results and empirical data also reveal power-law asymptotic behavior of burst duration, peak value and burst size scatter plots. The exponents of observed  power-law statistics probably have universal nature and will be considered in our future work.

Our analysis of empirical data suggests that long range fluctuations of absolute return in financial markets can be modeled by non-linear stochastic differential equations with $\eta$ up to $\frac{5}{2}$.

\section*{Acknowledgments}
We would like to thank Dr. Julius Ruseckas for valuable comments on the first draft of this work. The authors acknowledge the support by the EU COST Action MP0801 Physics of Competition and Conflicts stimulating our international cooperation.

\appendix

\section{Double stochastic return model}
The class of equations based on SDE (\ref{eq:sde}) gives only a general idea how
to model power-law statistics of trading activity and return in the financial markets.
The problem is to determine the parameter set $\lambda$ and $\eta$, which would
enable reproduction of the empirical values of $\lambda$ and $\beta$.
The task becomes even more complicated if one considers the more sophisticated
trends in the spectral density - power spectral densities have not a
single, but two power-law regions with different values of $\beta$. In the series of
papers \cite{Gontis2006JStatMech,Gontis2008PhysA,Gontis2010PhysA} we have
shown that trading activity and return can be modeled by a more sophisticated
stochastic differential equation than (\ref{eq:sde}) now including the two powers of the noise
multiplicativity. In the case of return instead of Eq. (\ref{eq:sde}) one should
use \cite{Gontis2010PhysA}
\begin{equation}
\rmd x=\left[\eta-\frac{\lambda}{2}-\left(\frac{x}{x_{max}}\right)^2\right]
\frac{(1+x^2)^{\eta-1}}{(\epsilon \sqrt{1+x^2}+1)^2}x\rmd
t_s + \frac{(1+x^2)^{\frac{\eta}{2}}}{\epsilon \sqrt{1+x^2}+1}\rmd
W_s, \label{eq:return2}
\end{equation}
here $\epsilon$ divides the area of diffusion into the two different
noise multiplicativity regions to ensure the spectral density of $|x|$ with two
power law exponents, term $\left(\frac{x}{x_{max}}\right)^2$ gives the exponential diffusion restriction for
large values of variable when $x>x_{max}$.

The proposed form of the more complex SDE enables reproduction of the main statistical
properties of the return observed in the financial markets.
This provides an approach to the financial markets with behavior dependent on
the level of activity and exhibiting two stages: calm and excited.
One more peculiarity of the proposed model is that signal $x$ serves only as modulating one of
secondary high frequency $q$-Gaussian fluctuations.  We formalized the empirical return
$r_t$ in the model as instantaneous q-Gaussian fluctuations $\xi$ with a slowly diffusing
parameter $r_0$ and constant $\lambda=5$
\begin{equation} \label{eq:firstModel}
r_t = \xi \{ r_0, \lambda \} .
\end{equation}
q-Gaussian distribution of $\xi$ can be expressed as follows:
\begin{equation}
P_{r_0,\lambda} (r) = \frac{\Gamma\left({\frac{\lambda}{2}}\right)}{r_0 \sqrt{\pi} \Gamma\left({\frac{\lambda}{2}-\frac{1}{2}}\right)} \left (\frac{r_0^2}{r_0^2+r^2} \right)^{\frac{\lambda}{2}} ,
\end{equation}
The variable $r_0$ serves as a measure of instantaneous volatility of high frequency return
fluctuations \cite{Gontis2010PhysA}. We proposed to model the measure of volatility $r_0$
by the scaled continuous stochastic variable $x$, which can be interpreted as average
return per unit time interval. Through the empirical analysis of high frequency empirical data
from NYSE \cite{Gontis2010PhysA} we have introduced the following relation
\begin{equation}
r_0 (t,\tau)=1+\frac{\bar r_0}{\tau_s} \left | \int\limits_{t_s}^{t_s+\tau_s} x(s) \mathrm{d} s \right | ,
\end{equation}
where $\bar{r}_0$ is an empirical parameter and the average return per unit time interval
$x(t_s)$ can be modeled by a nonlinear SDE (\ref{eq:return2}), expressed in scaled
dimensionless time $t_s = \sigma_t^2 t$.

\bibliographystyle{ieeetr}
\bibliography{gontis-kononovicius-ws-acs}

\begin{thebibliography}{10}

\bibitem{jeanblanc2009}
M.~Jeanblanc, M.~Yor, and M.~Chesney, {\em Mathematical Methods for Financial
  Markets}.
\newblock Berlin: Springer, 2009.

\bibitem{Davydov2001}
D.~Davydov and V.~Linetsky, ``Pricing and hedging path-dependent options under
  the cev process,'' {\em Management Science}, vol.~47, pp.~949--965, 2001.

\bibitem{Chan1992}
K.~C. Chan, G.~Andrew~Karolyi, F.~A. Longstaff, and A.~B. Sanders, ``An
  empirical comparison of alternative models of the short-term interest rate,''
  {\em THE JOURNAL OF FINANCE}, vol.~XLVII, no.~3, pp.~1209--1227, 1992.

\bibitem{Gontis2004PhysA343}
V.~Gontis and B.~Kaulakys, ``Multiplicative point process as a model of trading
  activity,'' {\em Physica A}, vol.~343, pp.~505--514, 2004.

\bibitem{Kaulakys2005PhysRev}
B.~Kaulakys, V.~Gontis, and M.~Alaburda, ``Point process model of 1/f noise vs
  a sum of lorentzians,'' {\em Phys. Rev. E}, vol.~71, no.~051105, pp.~1--11,
  2005.

\bibitem{Ruseckas2011EPL}
J.~Ruseckas, B.~Kaulakys, and V.~Gontis, ``Herding model and 1/f noise,'' {\em
  EPL}, vol.~96, p.~60007, 2011.

\bibitem{Gontis2010Sciyo}
V.~Gontis, J.~Ruseckas, and A.~Kononovicius, {\em A Non-Linear Double
  Stochastic Model of Return in Financial Markets}, pp.~559--580.
\newblock No.~ISBN: 978-953-307-121-3, Sciyo, August 2010.

\bibitem{Gontis2010PhysA}
V.~Gontis, J.~Ruseckas, and A.~Kononovičius, ``A long-range memory stochastic
  model of the return in financial markets,'' {\em Physica A}, vol.~389,
  pp.~100--106, 2010.

\bibitem{gardiner1997}
C.~W. Gardiner, {\em Handbook of stochastic methods}.
\newblock Berlin: Springer, 1997.

\bibitem{Kaulakys2006PhysA}
B.~Kaulakys, J.~Ruseckas, V.~Gontis, and M.~Alaburda, ``Nonlinear stochastic
  models of 1/f noise and power-law distributions,'' {\em Physica A}, vol.~365,
  pp.~217--221, 2006.

\bibitem{Ruseckas2010}
J.~Ruseckas and B.~Kaulakys, ``1/f noise from nonlinear stochastic differential
  equations,'' {\em Physical Review E}, vol.~81, p.~031105, 2010.

\bibitem{Gabaix2006}
X.~Gabaix, P.~Gopikrishnan, V.~Plerou, and H.~E. Stanley, ``Institutional
  investors and stock market volatility,'' {\em The Quarterly Journal of
  Economics}, pp.~461--504, 2006.

\bibitem{ReimmanPhysA2011}
S.~Reimann, V.~Gontis, and M.~Alaburda, ``Interplay between positive feedbacks
  in the generalized cev process,'' {\em Physica A: Statistical Mechanics and
  its Applications}, vol.~390, no.~8, pp.~1393--1401, 2011.

\bibitem{Daniunas2011}
V.~Daniunas, V.~Gontis, and A.~Kononovicius, ``Agent-based versus macroscopic
  modeling of competition and business processes in economics,'' {\em ICCGI
  2011 : The Sixth International Multi-Conference on Computing in the Global
  Information Technology}, pp.~84--88, 2011.

\bibitem{Kononovicius2012PhysA}
A.~Kononovicius and V.~Gontis, ``Agent based reasoning for the non-linear
  stochastic models of long-range memory,'' {\em Physica A}, vol.~391, no.~4,
  pp.~1309--1314, 2012.

\bibitem{Kirman1993}
A.~P. Kirman, ``Ants, rationality and recruitment,'' {\em Quarterly Journal of
  Economics}, vol.~108, pp.~137--156, 1993.

\bibitem{Alfarano2005}
S.~Alfarano, T.~Lux, and F.~Wagner, ``Estimation of agent-based models: The
  case of an asymmetric herding model,'' {\em Computational Economics},
  vol.~26, no.~1, pp.~19--49, 2005.

\bibitem{Alfarano2008}
S.~Alfarano, T.~Lux, and F.~Wagner, ``Time variation of higher moments in a
  financial market with heterogeneous agents: An analytical approach,'' {\em
  Journal of Economic Dynamics and Control}, vol.~32, pp.~101--136, 2008.

\bibitem{redner2001}
S.~Redner, {\em A guide to first-passage processes}.
\newblock Cambridge University Press, 2001.

\bibitem{Kloeden1999Springer}
P.~E. Kloeden and E.~Platen, {\em Numerical Solution of Stochastic Differential
  Equations}.
\newblock Berlin: Springer, 1999.

\bibitem{Borodin2002Birkhauser}
A.~N. Borodin and P.~Salminen, {\em Handbook of Brownian Motion}.
\newblock Basel, Switzerland: Birkhauser, 2~ed., 2002.

\bibitem{Abramowitz1972}
M.~Abramowitz and I.~A. Stegun, {\em Handbook of Mathematical Functions with
  Formulas, Graphs, and Mathematical Tables}.
\newblock New York: Dover, 1972.

\bibitem{Gontis2006JStatMech}
V.~Gontis and B.~Kaulakys, ``Long-range memory model of trading activity and
  volatility,'' {\em J. Stat. Mech.}, vol.~P10016, pp.~1--11, 2006.

\bibitem{Gontis2008PhysA}
V.~Gontis, B.~Kaulakys, and J.~Ruseckas, ``Trading activity as driven poisson
  process: comparison with empirical data,'' {\em Physica A}, vol.~387,
  pp.~3891--3896, 2008.

\end{thebibliography}

\end{document}